# A FLOSS Visual EM Simulator for 3D Antennas

Christos A. Koutsos, Nikolitsa I. Yannopoulou, and Petros E. Zimourtopoulos

***Abstract*—This paper introduces the FLOSS Free Libre Open Source Software [VEMSA3D], a contraction of "Visual Electromagnetic Simulator for 3D Antennas", which are geometrically modeled, either exactly or approximately, as thin wire polygonal structures; presents its GUI Graphical User Interface capabilities, in interactive mode and/or in handling suitable formed antenna data files; demonstrates the effectiveness of its use in a number of practical antenna applications, with direct comparison to experimental measurements and other freeware results; and provides the inexperienced user with a specific list of instructions to successfully build the given source code by using only freely available IDE Integrated Development Environment tools—including a cross-platform one. The unrestricted access to source code, beyond the ability for immediate software improvement, offers to independent users and volunteer groups an expandable, in any way, visual antenna simulator, for a genuine research and development work in the field of antennas, adaptable to their needs.**

## I. INTRODUCTION

A LOT of amazing visual EM software simulators, both commercial and freeware, exist for many years now. However, to the best of authors' knowledge, none of these simulators has been ever released under an Open Source license—paid, free gratis or free libre. This situation, as it was discussed recently, precludes the independent users and volunteer groups with limited resources, like the authors' nonprofit group, from the genuine, state-of-the-art, scientific research, since, neither the code improvement, nor its adaptation to specific needs, is possible [1].

For that reason, the authors decided to develop and release under the GNU public license their own Visual Electromagnetic Simulator for 3D Antennas [VEMSA3D], although it certainly is a less well-equipped software application for the moment—but still a fully expandable one. That release is now perfectly permissible, because in 2005 NASA released in the public domain the FORTRAN source code of the well-known MoM Method of Moments Thin-Wire Computer Program, by J. H. Richmond [2], [3]. This is exactly the code on which the authors' group based its freeware simulators: [RichWire], a CLI Command Line Interface, and [DA], a MS Quick-Win FORTRAN derivative of it. These two simulators are under uninterrupted revision, improvement, expansion, and re-development, since a long time ago [1]. Therefore, the [RichWire] FORTRAN code was translated—entirely, line-by-line, without using any paid or free translator—to C++, to form the core of scientific EM computations in [VEMSA3D].

On the other hand, the authors' group requirements for scientific accuracy in visual representation of the produced EM simulation antenna results from its simulators, were already enforced the software development of the Virtual Antennas, that is the Virtual Antennas laboratories, in VRML [4], the FLOSS application [RadPat4W] for antenna radiation pattern presentation, in MS VB6 [1], [5], as well as, the recently developed visual antenna application for the Wolfram Demonstrations Project in Mathematica [6]. The visualization ideas implemented in the afore-mentioned graphics applications were also expressed—from the scratch—in C++, to use the Open Source cross-platform [wxWidgets] library with OpenGL and form the core of scientific EM graphics in [VEMSA3D] [7].

The authors, having taken into account that no familiarization with software use is possible without getting hands-on experience, restrict themselves to a brief discussion of the current [VEMSA3D] characteristics. The code, the antenna applications data, as well as, any other information, referenced open bugs or future code releases, will be always available in authors' group repositories, at http://www.antennas.gr/floss and in GoogleCode website, at http://code.google.com/p/rga/ .

## II. GUI INTERACTIVE MODE OF OPERATION

It is assumed that the user has already some experience in the sketching of a polygonal wire outline model, for an antenna under consideration, consisting of numbered wire segments and nodes, with their 3D space coordinates, including the positions of any antenna circuit elements, that is voltage generators and lumped loads, and s/he wants then to key-in these model and circuit data into [VEMSA3D] using its GUI Graphical User Interface in interactive mode.

The GUI main window is shown in Fig. 1, while Fig. 2 shows the menu items along with their available submenu options numbered from 1 to 7.

The GUI itself is divided in three panels named [Antenna in space (3D)], [Antenna Elements], and [Datasets]. The function of each panel is briefly described in the following.

C. A. Koutsos is with Antennas Research Group, Hlavné námestie 4, 811 01 Bratislava, Slovak Republic (e-mail: cak@antennas.gr).

N. I. Yannopoulou is with Antennas Research Group, Palaia Morsini AO AA, PO Box 63, Xanthi, Thrace, Greece (e-mail: yin@antennas.gr).

P. E. Zimourtopoulos is with nonprofit Antennas Research Group, Palaia Morsini, Xanthi, Thrace, EU, Deptartment of Electrical Engineering and Computer Engineering, Democritus University of Thrace, V. Sofias 12, 671 00 Xanthi, Greece (e-mail: pez@antennas.gr).

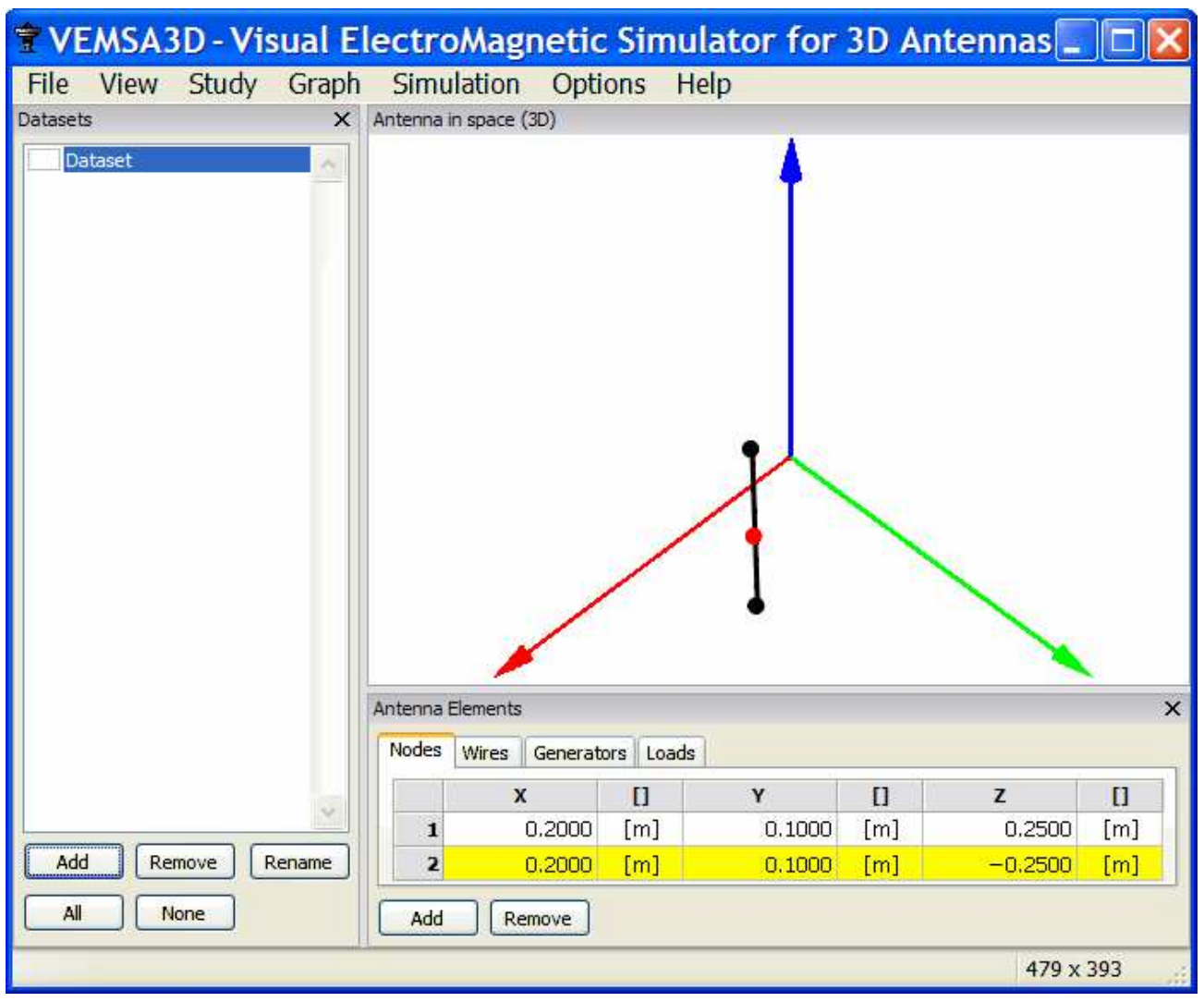

Fig. 1. GUI: The main window.

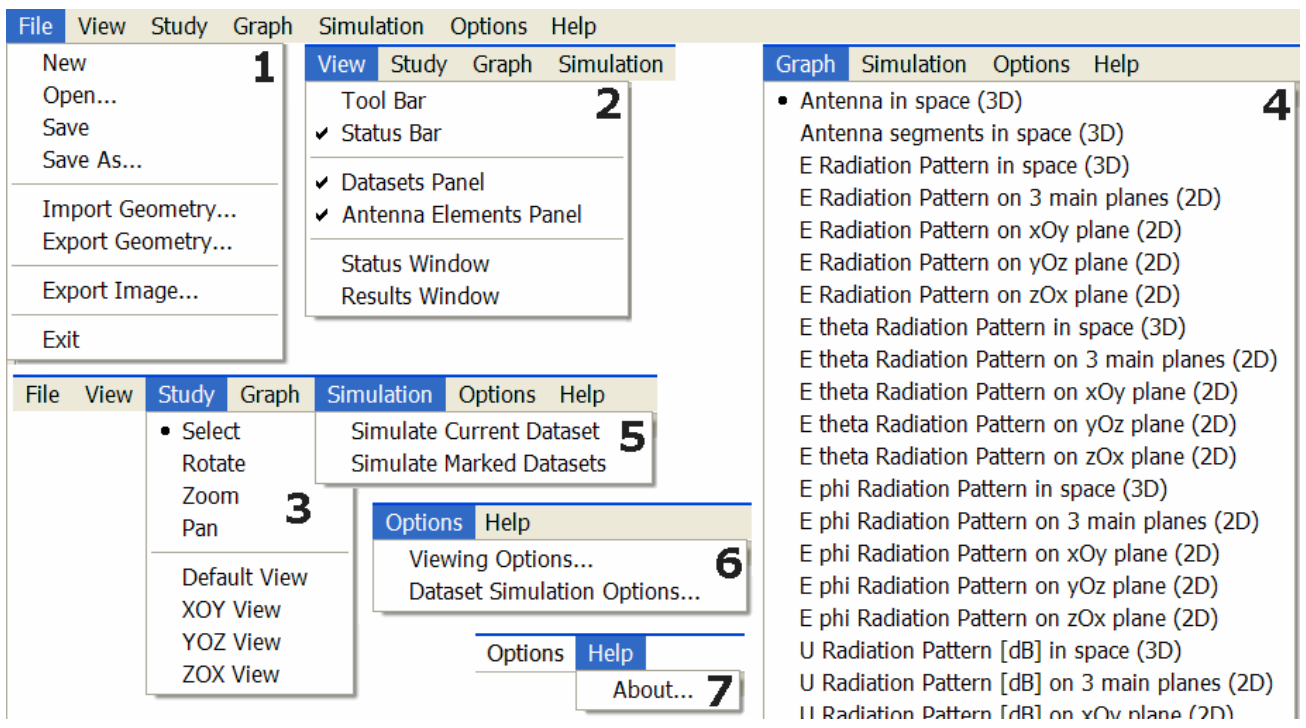

Fig. 2. Main window: The unfolded menu items.

The [Antenna in space (3D)] viewing panel is used to project all the generated 3D and 2D graphics. By default, the application starts with a simple linear dipole antenna in space, as it is shown in Fig. 1 from a viewpoint on the diagonal of trihedral angle OXYZ, where the usual letters associated with the Cartesian axes do not exist. Instead, a one-to-one correspondence implicitly exists between (X, Y, Z) axes and (R, G, B) colors [4]. The 3D image of any antenna can be manipulated through the [Study] options, as shown in Fig. 2.3.

The [Antenna Elements] panel contains four tabs, corresponding to model and circuit data, in four self-explained building blocks. Instead of attempting to describe the block data in general the presentation continues with the default data:

[Nodes]: 2, with coordinates 1: (0.2, 0.1, 0.25), 2: (0.2, 0.1, –0.25), in meters [m] or in terms of wavelength λ [wl].

[Wires]: 1, starting at node: 1, ending at node: 2, that is a directed segment. [Segments] is either: (a) a positive integer, (b) a zero or (c) a number between 0.00 and 0.25, to respectively define the division of this wire in segments: (a) in the indicated number of segments, (b) in a calculated number of segments of length no longer than 0.05λ (that is 10 segments, in this case), (c) in segments with length no more to that number. Notably, no segment can be longer than 0.25λ [2].

[Generators]: 1, connected in the wire:1, at a distance of 0.5 times this wire length, from its starting node, with rms value of 1∠0° [V]. There can be only one generator on each wire.

[Loads]: Has the same structure as [Generators]. Resistors, inductors and capacitors are inputted in [Ω], [H] and [F], respectively. It is empty in this case, since no lamped loads exist.

Notably, the GUI has been developed in such a way that efficiently supports the interactive handling of the antenna modeling by the user on the screen. Thus, the selection of an element in any of these four tabs simultaneously highlights this element on the viewing [Antenna in space (3D)] panel—if the [Select] option under [Study] (Fig. 2.3) has been already chosen by the user, and this is the default selection in any case—and conversely: the selection of any element on this viewing [Antenna in space (3D)] panel highlights the element in its table. Consequently, a full control on the antenna modeling structure is achieved.

The [Datasets] panel may contain independent sets of model and circuit data for various antennas. Such a Dataset has its own [Dataset Simulation Options], which are chosen through submenu of Fig. 2.6, as the default selection of them is shown on the left part of Fig. 3.

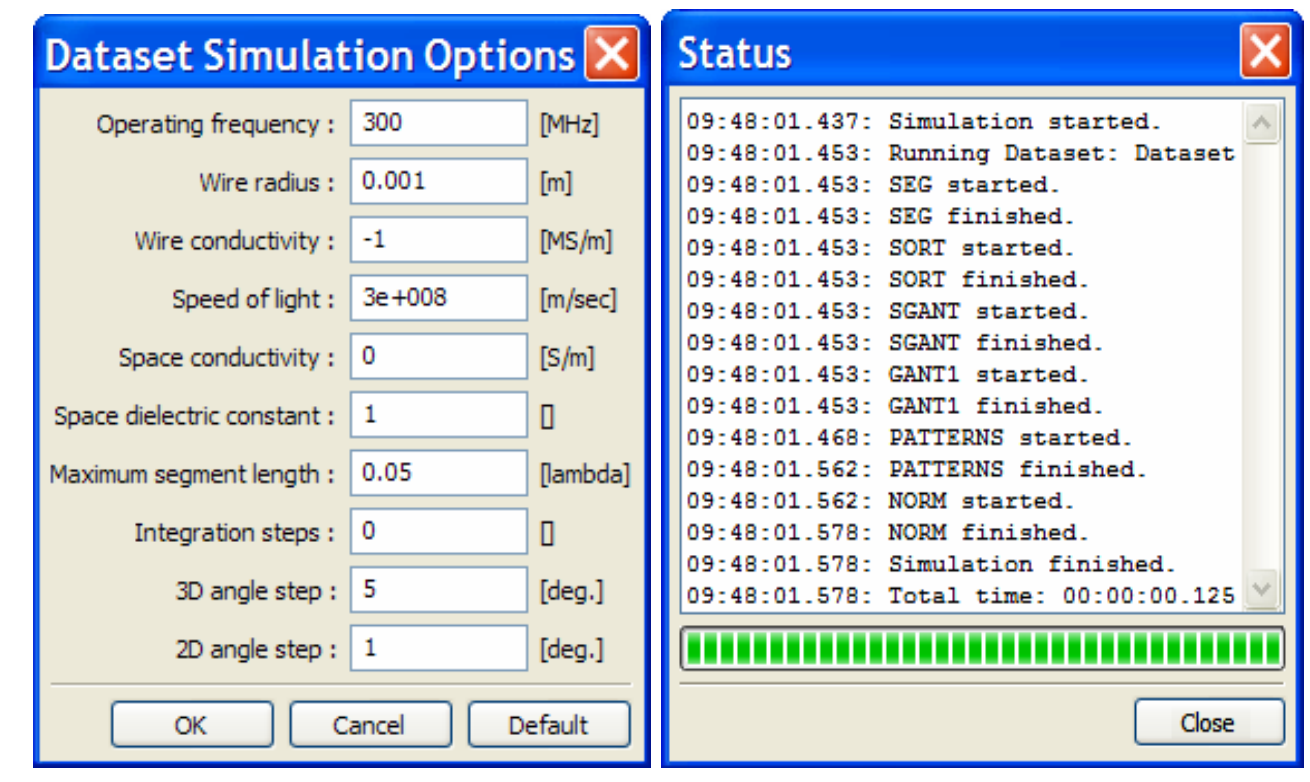

Fig. 3. Windows: [Dataset Simulation Options] - [Status].

A [Wire conductivity] of –1 corresponds to perfect wire conductivity, while the [Space conductivity] of 0 and the [Space dielectric constant] of 1 defines the Free Space EM environment. The [Integration steps] accepts a positive number for the steps of approximated integration used in the MoM impedance computations—a zero means closed-form integration.

Notably, the [Maximum segment length] affects the division of wires: the smaller this number is, the more segments will be considered. It is currently known that there is an implied, still programmatically unimproved, priority of the number of segments defined in [Wires] tab, over this selection—that is a bug, which may crash the application.

The EM simulation of an antenna starts by selecting [Simulate Current Dataset], under [Simulation] (Fig. 2.5), while its progress is shown in the [Status] window, such as the one at the right of Fig. 3, which concerns the default dipole. Notably, multiple Datasets can be simulated, one after the other, through the [Simulate Marked Datasets] shown in Fig. 2.5.

The course of the simulation may be roughly described in five steps: (1) Wires are automatically segmented, using the provided parameters, to produce the final model structure of points and segments - (2) Model structure is analyzed and adjacent segments are combined to form dipole current modes with sinusoidal distribution - (3) Impedances are calculated and the MoM algebraic system of equations is formed - (4) The system of equations is solved and the segment currents are calculated - (5) 3D and 2D radiation patterns are calculated, as well as other useful results like those shown in Fig. 4.

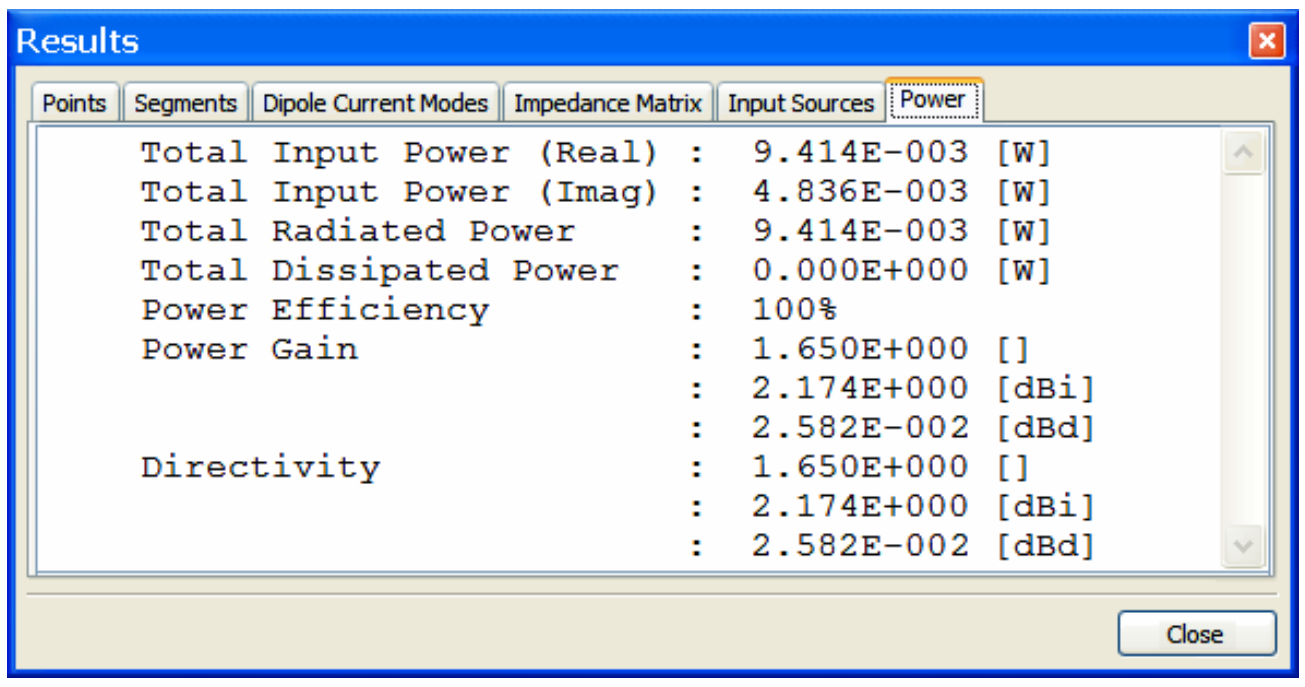
Results

Points | Segments | Dipole Current Modes | Impedance Matrix | Input Sources | Power

```
Total Input Power (Real)  :  9.414E-003 [W]
Total Input Power (Imag)  :  4.836E-003 [W]
Total Radiated Power      :  9.414E-003 [W]
Total Dissipated Power    :  0.000E+000 [W]
Power Efficiency          :  100%
Power Gain                :  1.650E+000 []
                          :  2.174E+000 [dBi]
                          :  2.582E-002 [dBd]
Directivity               :  1.650E+000 []
                          :  2.174E+000 [dBi]
                          :  2.582E-002 [dBd]
```

Close

Fig. 4. Window: [Results] - Table: [Power].

In Fig. 5, the [Viewing Options] under the [Options] (Fig. 2.6) is given with all of the available options for the graphics.

Viewing Options

| | | | | | |
|---|---|---|---|---|---|
| Show Axis : | ☑ | Axis Size : | 15 | Background Color : | #FFFFFF |
| Show Nodes/Points : | ☑ | Antenna Scale : | 500 | X Axis Color : | #FF0000 |
| Show Wires/Segments : | ☑ | Nodes/Points Size : | 50 | Y Axis Color : | #00FF00 |
| Show Generators : | ☑ | Wires/Segments Size : | 15 | Z Axis Color : | #0000FF |
| Show Loads : | ☑ | Generators Size : | 100 | Nodes/Points Color : | #000000 |
| Show Current Amplitude : | ☐ | Loads Size : | 100 | Wires/Segments Color : | #000000 |
| Show Current Phase : | ☐ | Current Amplitude Line Width : | 20 | Generators Color : | #FF0000 |
| Show Maps : | ☑ | Current Amplitude Size : | 30 | Loads Color : | #00FF00 |
| Show 3D Radiation Pattern : | ☑ | Current Phase Line Width : | 20 | Current Amplitude Color : | #FF00FF |
| Show Radiation Pattern Planes 1 : | ☑ | Current Phase Size : | 30 | Current Phase Color : | #00FFFF |
| Show Radiation Pattern Planes 2 : | ☑ | Maps Size : | 10 | Maps Color : | #000000 |
| Show 2D Radiation Patterns : | ☐ | Maps dB Depth : | -20 | Radiation Pattern Color : | #858585 |
| | | Radiation Pattern 2D Line Width : | 40 | | |

Default

Fig. 5. [Viewing Options].

Through [Graph] menu item (Fig. 2.4), a variety of 3D and 2D plots, which include either the normalized electric far field E radiation pattern or the relative radiation intensity U in dB, as well as their θ and φ parts, can be illustrated.

The [Antenna segments in space (3D)] illustrates the final antenna structure with points, segments and segment currents in amplitude and phase, as shown in Fig. 6, where the resulting current distribution on default dipole it seems to be sinusoidal, that is as it was expected to be. The well-known 2D intersections of the 3D radiation patterns by the three main planes yOz, xOy, zOx, as well as these 3D patterns themselves are also shown for the default dipole.

Notably, modeled antennas can be imported and exported in [RichWire] data file format (Fig 2.1).

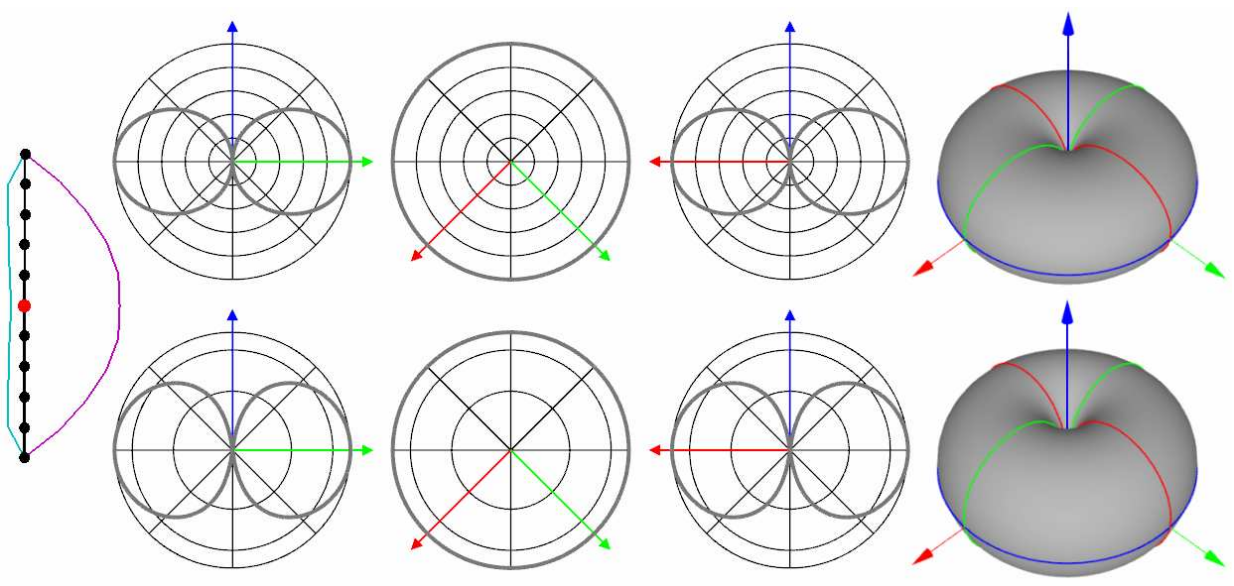

Fig. 6. GUI: Default dipole results.

## III. BUILD UNDER MS WINDOWS XP

To build [VESMA3D] MS Windows executables, the use of IDE CodeBlocks with GCC/Mingw is suggested [8]:

```
Set below as X:\ the system disk, e.g. X:\ -> C:\

1 Download (~20MB), Install (~110MB):
  http://sourceforge.net/projects/codeblocks
 [View all files][Binaries][8.02]
  codeblocks-8.02mingw-setup.exe
  Select: [Full], [Run], Compiler: [GCC/Mingw]

2 Download wxWidgets (~20MB), Extract (~120MB):
  http://sourceforge.net/projects/wxwindows
 [View all files] wxWidgets-2.8.11.zip
  wxWidgets-2.8.11.zip-> X:\ [Yes to All]

3 IDE CodeBlocks, Open:
  X:\wxWidgets-2.8.11\include\wx\msw\setup0.h
  Replace, at line 1006:
 #define wxUSE_GLCANVAS 0 -> 1
  Repeat it, to file: setup.h

4 At system variable [path]:
 [My Computer][Properties][Advanced]
 [Environment Variables][System Variables]
  Path[Edit][Variable value], Add:
 ";X:\Program Files\CodeBlocks\MinGW\bin;"

5 Command Prompt: >cd\
 >cd X:\wxWidgets-2.8.11\build\msw
 >mingw32-make -f makefile.gcc MONOLITHIC=0
  SHARED=0 UNICODE=0 USE_OPENGL=1
  BUILD=release

  Re-Command, with option:
  BUILD=release -> BUILD=debug

6 Download (~100KB), Extract (~700KB):
  http://code.google.com/p/rga/downloads/
  VEMSA3D_source_1.zip -> X:\

7 IDE CodeBlocks, [File][Open]:
  X:\VEMSA3D\build\win-cb\VEMSA3D.workspace

   * If system disk is not C:\ then:
 (#)[Project][Build options...][VEMSA3D]
   * [Search directories] Correct in each of:
   * [Compiler],[Linker],[Resource compiler]
   * C:\ -> X:\ [Yes]
   * Repeat from (#), for [Release]
   * Repeat from (#), for [Debug]

 [Build][Select target][Release][Build]
 [Build][Select target][Debug][Build]
```

This process results in [VEMSA3D] executables of ~35MB Debug and of ~5MB Release versions, which run under NT4, W2K and WXP, at least.

Alternatively, the use of the freeware IDE MS Visual Studio Express with C++ 2008 Compiler is also suggested. For that, first download MS Visual Studio 2008 Express iso-image from Microsoft website, burn it into a CD-R, and install it. Then, download from our repositories the file [VEMSA3Dfiles4win-cb.zip], extract it, and follow the included setup instructions. This process results in [VEMSA3D] executables of ~5MB Debug and of ~2MB Release versions, which run under W2K and WXP, at least, but definitely do not run under NT4.

## IV. Practical Antenna Applications and Results

This section presents the implementation of [VEMSA3D] to produce eight antenna models, from simple to more complicated structures. The input data were imported through [Import Geometry] of the [File] menu (Fig. 2.1).

Fig. 7 shows an array of 2 dipoles for operation at the frequency of 1111 MHz, distanced by 0.85λ, constructed by bare copper wire of 1 mm (0.0037λ) radius, and measured. In the same figure, the 3D radiation intensity pattern and its 2D main plane cuts are shown [9]. The continuous line represents [VEMSA3D] results, the dashed line, analytically produced patterns, and the dots, measurements made using the authors' group VNA system [10].

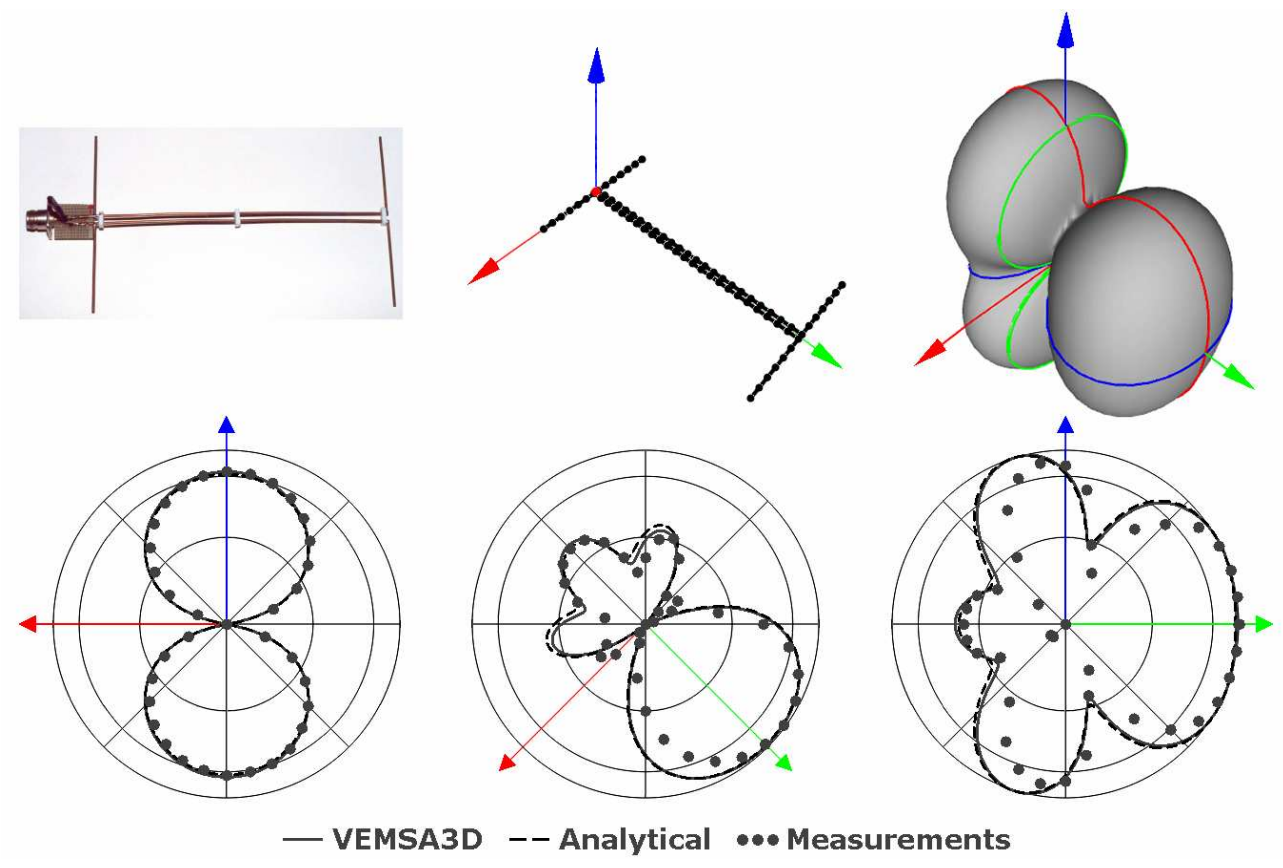

Fig. 7. Array of 2 dipoles: 73 dipole current modes.

Fig. 8 illustrates the results for a constructed and measured improved Hentenna model, at 1110 MHz, with height of λ/2, width of λ/6, and with the active element at a distance 7λ/12 from its bottom [11].

The third antenna consists of a λ/4 monopole (5.83 cm) at 1286 MHz over a circular counterpoise of 14 cm radius. The monopole was constructed by 2 mm diameter copper wire and the counterpoise has been printed on a circuit board of 29 cm x 29 cm, as a circle with four radials of 2 mm width [12]. In Fig. 9, the left 3D and 2D patterns are for vertical polarization, while the right ones for horizontal. For a better representation, the patterns have been normalized.

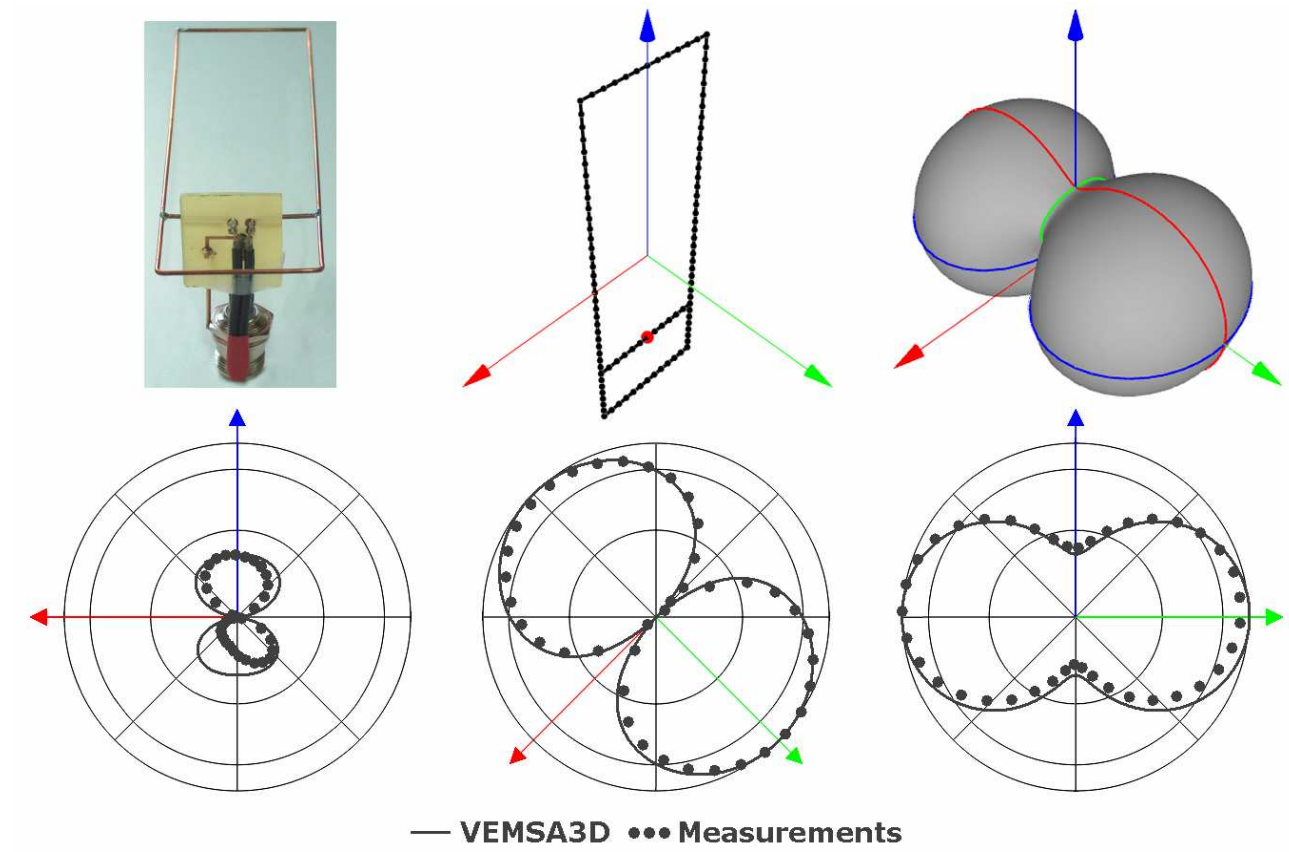

Fig. 8. Hentenna: 109 dipole current modes.

A somehow more complicated antenna is shown in Fig. 10. The left 3D and 2D patterns correspond to the vertical polarization of a monopole over circular counterpoise with 16 radials, while the right column patterns are the corresponding 3D and 2D patterns for the same antenna, but with an additional grounded disc below it, and without electrical connection to the counterpoise antenna, at a distance of 0.9 cm. The grounded disc was constructed from thin copper sheet and was modeled in [VEMSA3D] with 5 concentric circles and 64 radials [12].

In Fig. 11, the results of [VEMSA3D] for a commercial UHF antenna are shown [13]. The antenna model is presented in detail and separately, for the active element system, as well for the wire-frame reflector. The patterns correspond to the center frequency of its operation at 650 MHz.

Fig. 12 illustrates the model of a small jet airplane from the well-known freeware [4NEC2X] antenna simulator [14]. Simulation was carried out at the frequency of 10 MHz with the same number of points and segments in both simulators. There is a good agreement between them for the radiation intensity patterns although some deviation in input impedance and directivity is observed.

Fig. 13 shows the results for a horn antenna with dimensions proposed by K. Pitra and Z. Raida [15]. A small bow-tie feeder with a triangular perimeter of 0.47λ and flare angle 37.5°, is considered. The antenna was initially simulated at the frequency of 40 GHz in both [4NEC2X] and [VEMSA3D] simulators, with 1205 segments to be consistent with the restricted maximum number of 1500 segments of [4NEC2X]. There is a good agreement between the produced results.

Finally, the most complicated antenna model, which is presented, corresponds to the same horn antenna at the same frequency of 40 GHz, with a dense wire-frame consisting of 3266 points and 4362 segments. The model is shown in Fig. 14. The total number of dipole current modes is 5458, as shown in Fig. 15. The process time ranges from ~40 min in an AMD Phenom X2 550 3.11 GHz CPU to ~220 min, in an Intel Pentium 4 1.7 GHz CPU. Fig. 16 illustrates the resulting radiation patterns.

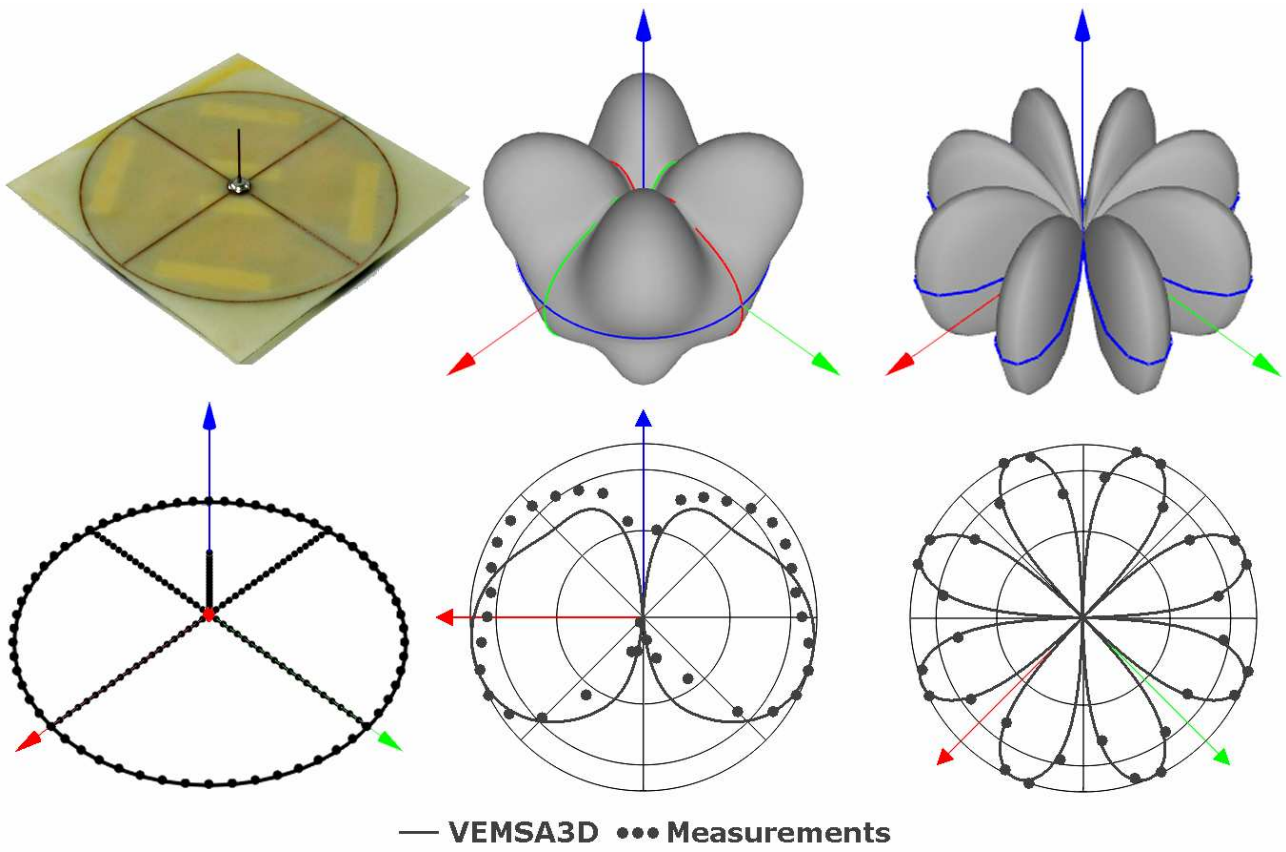

Fig. 9. Monopole over a counterpoise with 4 radials:
167 dipole current modes.

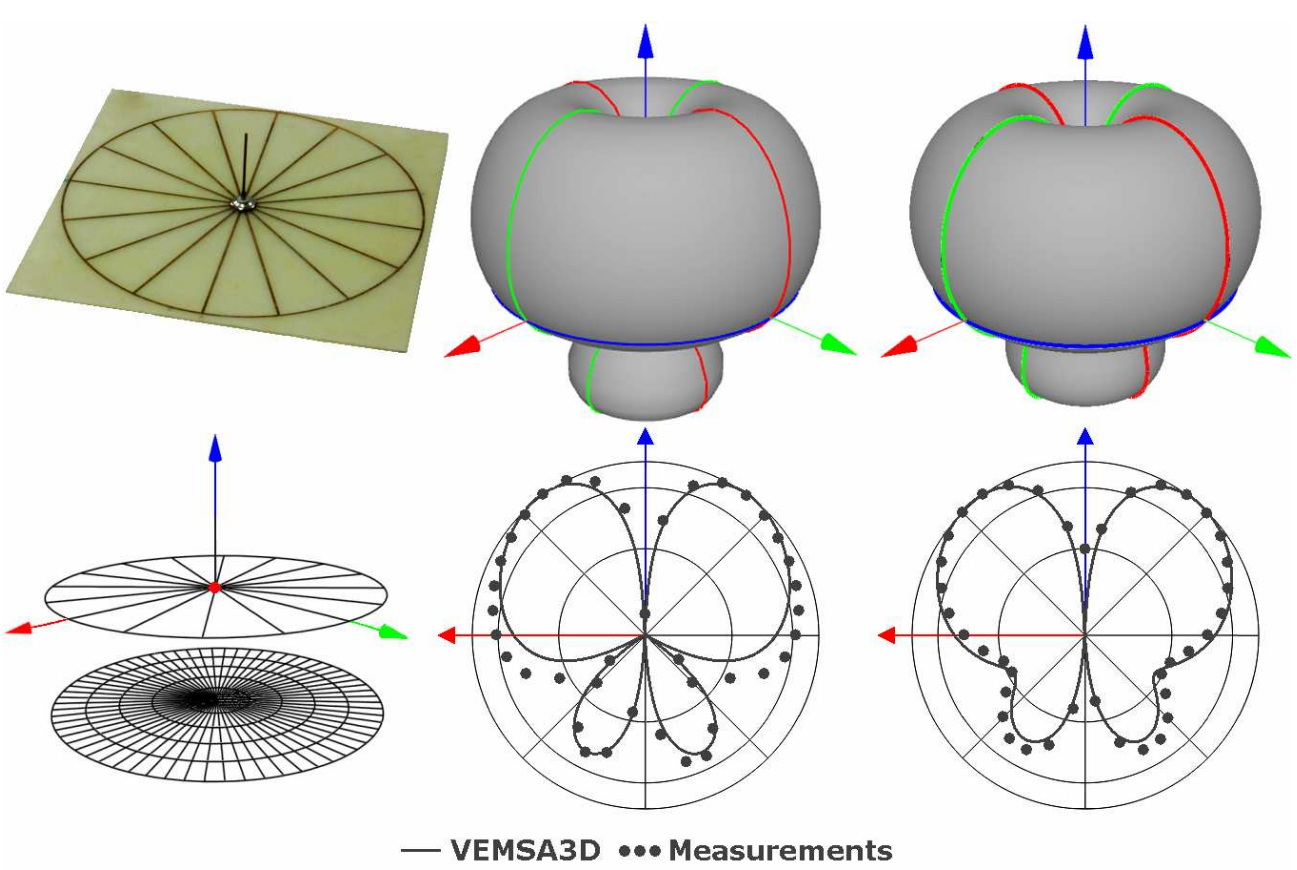

Fig. 10. Monopole over a counterpoise with 16 radials:
419 dipole current modes.
Monopole over a counterpoise with 16 radials over a grounded disc:
2594 dipole current modes.

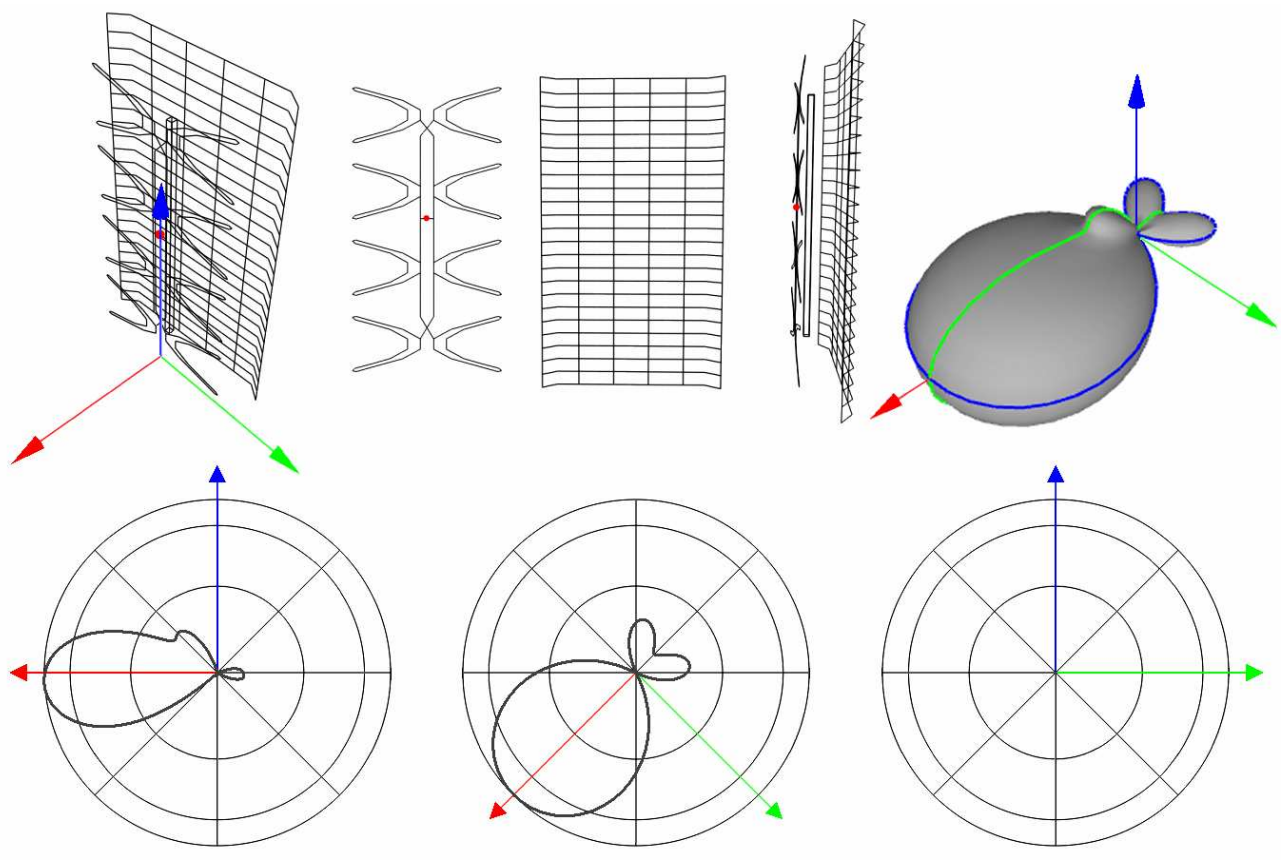

Fig. 11. Typical commercial TV UHF antenna:
723 dipole current modes.

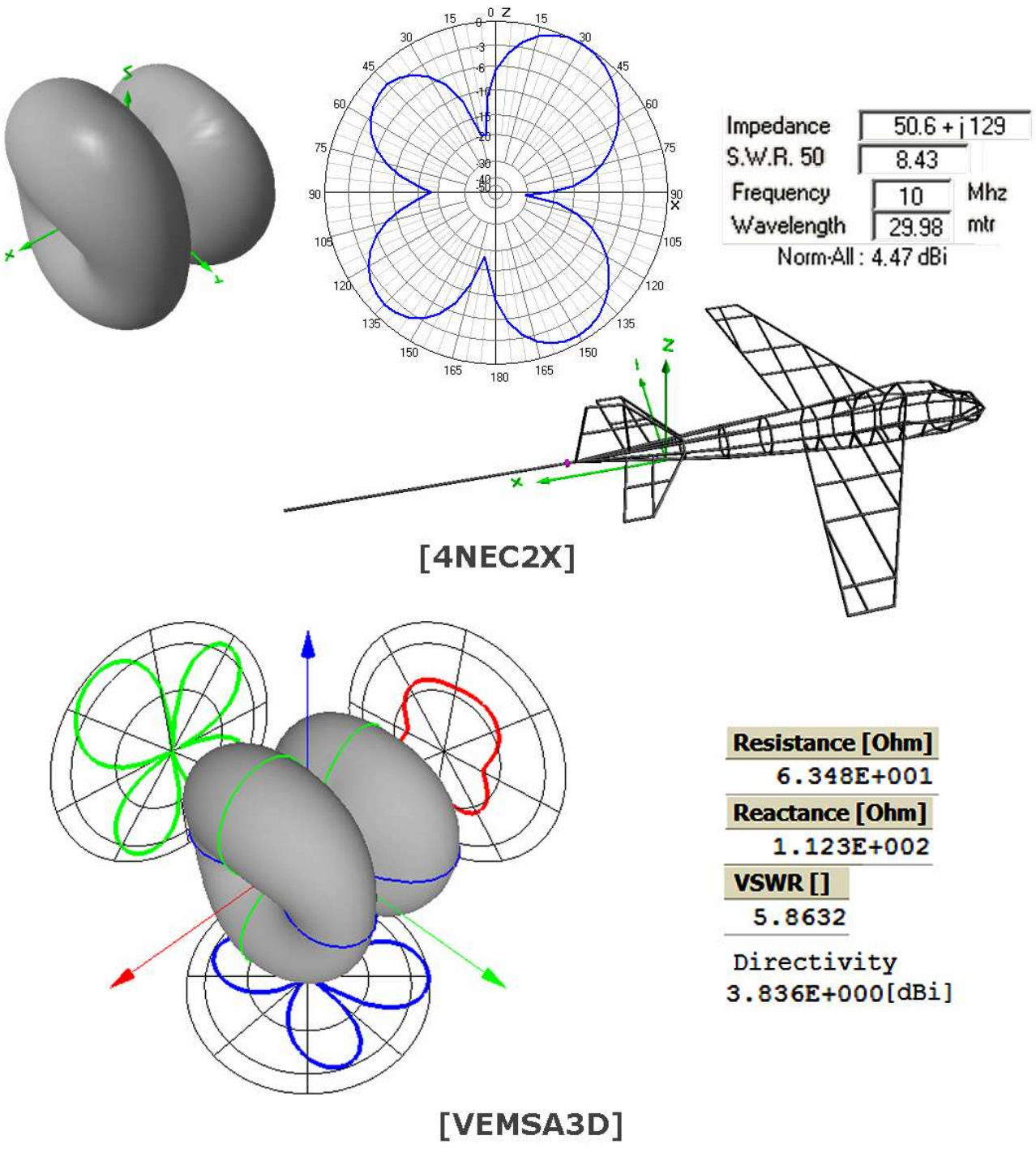

Fig. 12. Jet airplane modeled in
[4NEC2X] and [VEMSA3D]:
391 dipole current modes

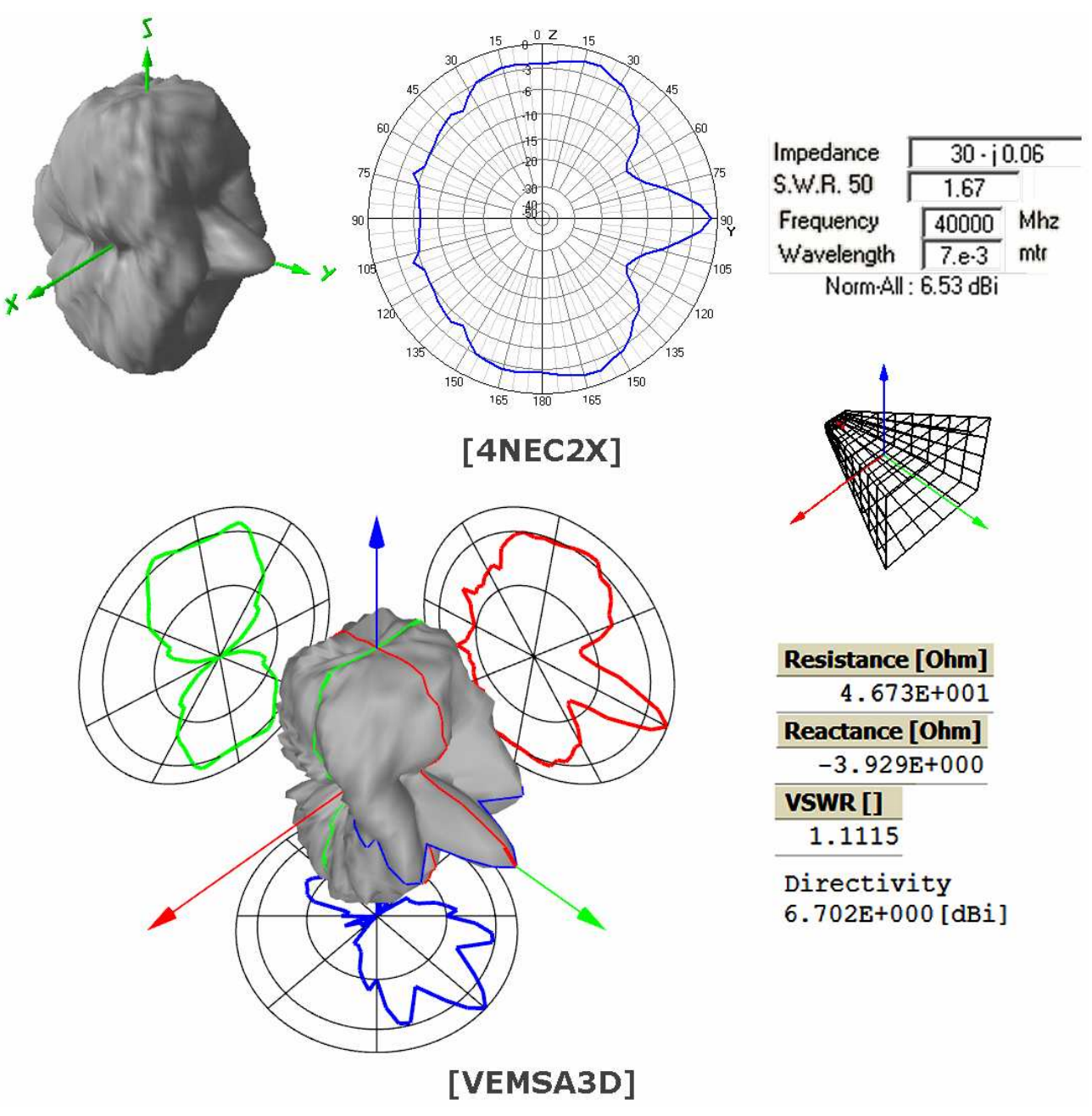

Fig. 13. Horn antenna with bow-tie feeder modeled in
[4NEC2X] and [VEMSA3D]:
1370 dipole current modes.

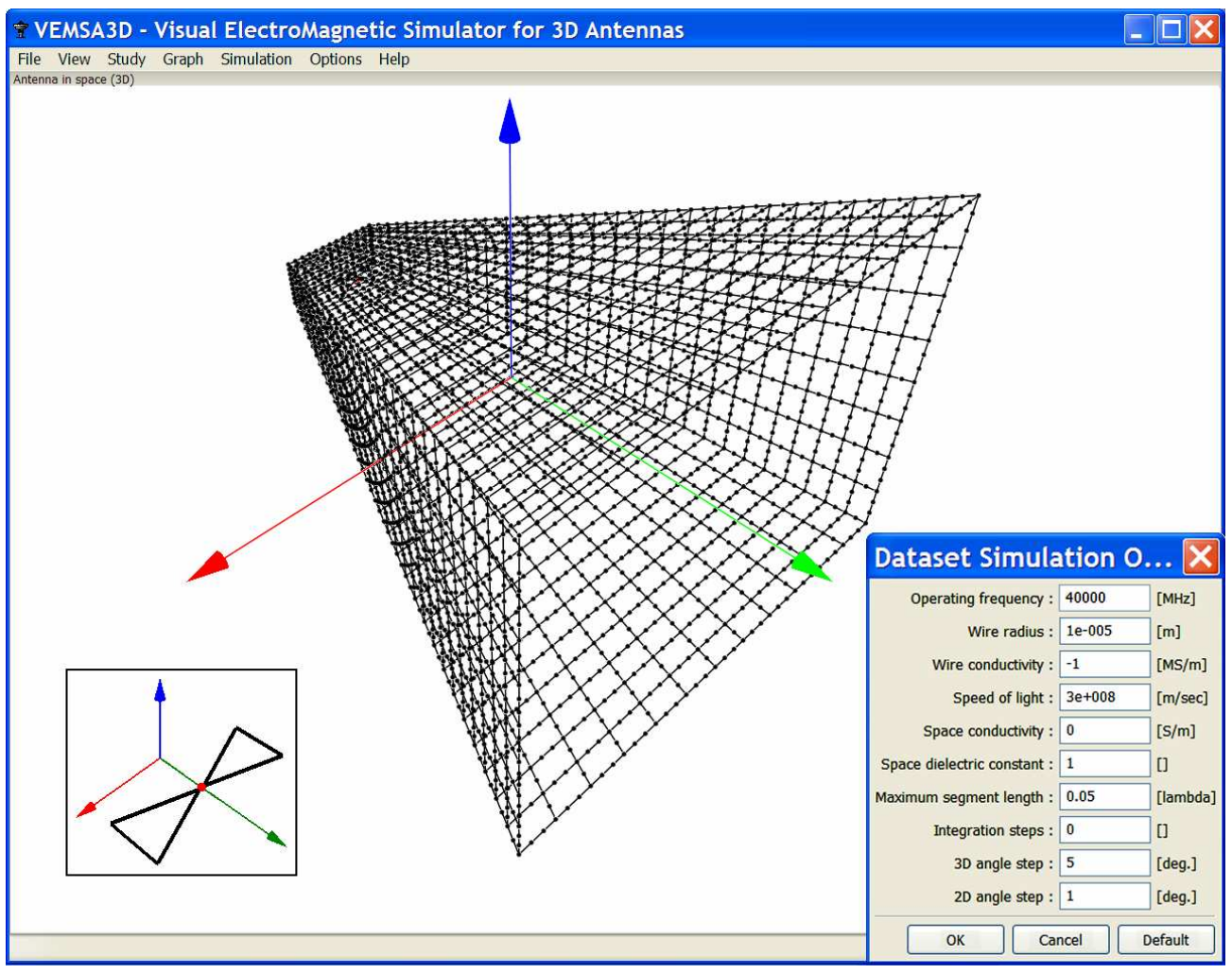

Fig. 14. Horn antenna with bow-tie feeder: 4362 segments.

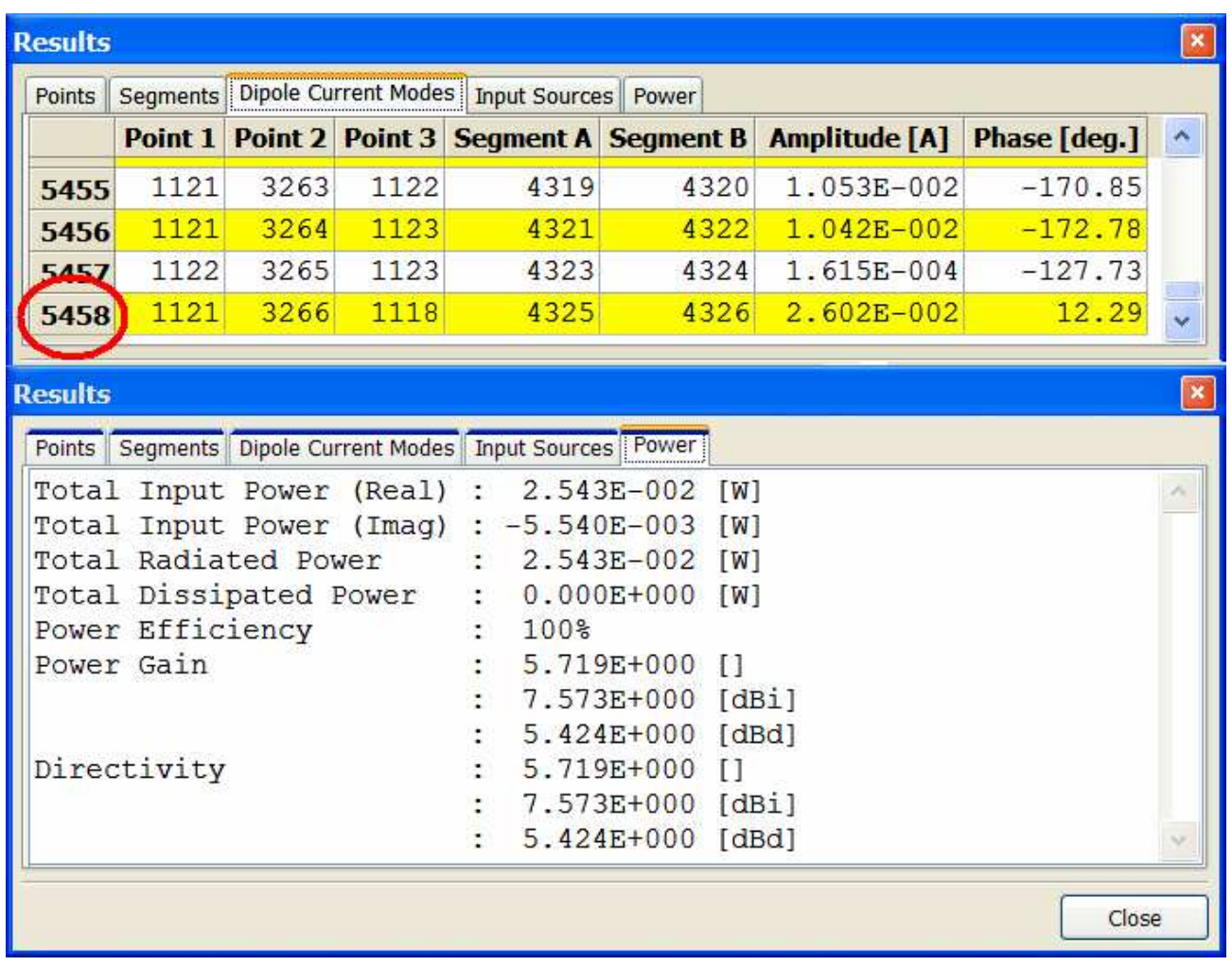

Results

Points | Segments | Dipole Current Modes | Input Sources | Power

| | Point 1 | Point 2 | Point 3 | Segment A | Segment B | Amplitude [A] | Phase [deg.] |
|---|---|---|---|---|---|---|---|
| 5455 | 1121 | 3263 | 1122 | 4319 | 4320 | 1.053E-002 | -170.85 |
| 5456 | 1121 | 3264 | 1123 | 4321 | 4322 | 1.042E-002 | -172.78 |
| 5457 | 1122 | 3265 | 1123 | 4323 | 4324 | 1.615E-004 | -127.73 |
| 5458 | 1121 | 3266 | 1118 | 4325 | 4326 | 2.602E-002 | 12.29 |

Results

Points | Segments | Dipole Current Modes | Input Sources | Power

```
Total Input Power (Real) :  2.543E-002 [W]
Total Input Power (Imag) : -5.540E-003 [W]
Total Radiated Power     :  2.543E-002 [W]
Total Dissipated Power   :  0.000E+000 [W]
Power Efficiency         :  100%
Power Gain               :  5.719E+000 []
                         :  7.573E+000 [dBi]
                         :  5.424E+000 [dBd]
Directivity              :  5.719E+000 []
                         :  7.573E+000 [dBi]
                         :  5.424E+000 [dBd]
```

Close

Fig. 15. Horn antenna with bow-tie feeder: 5458 dipole current modes.

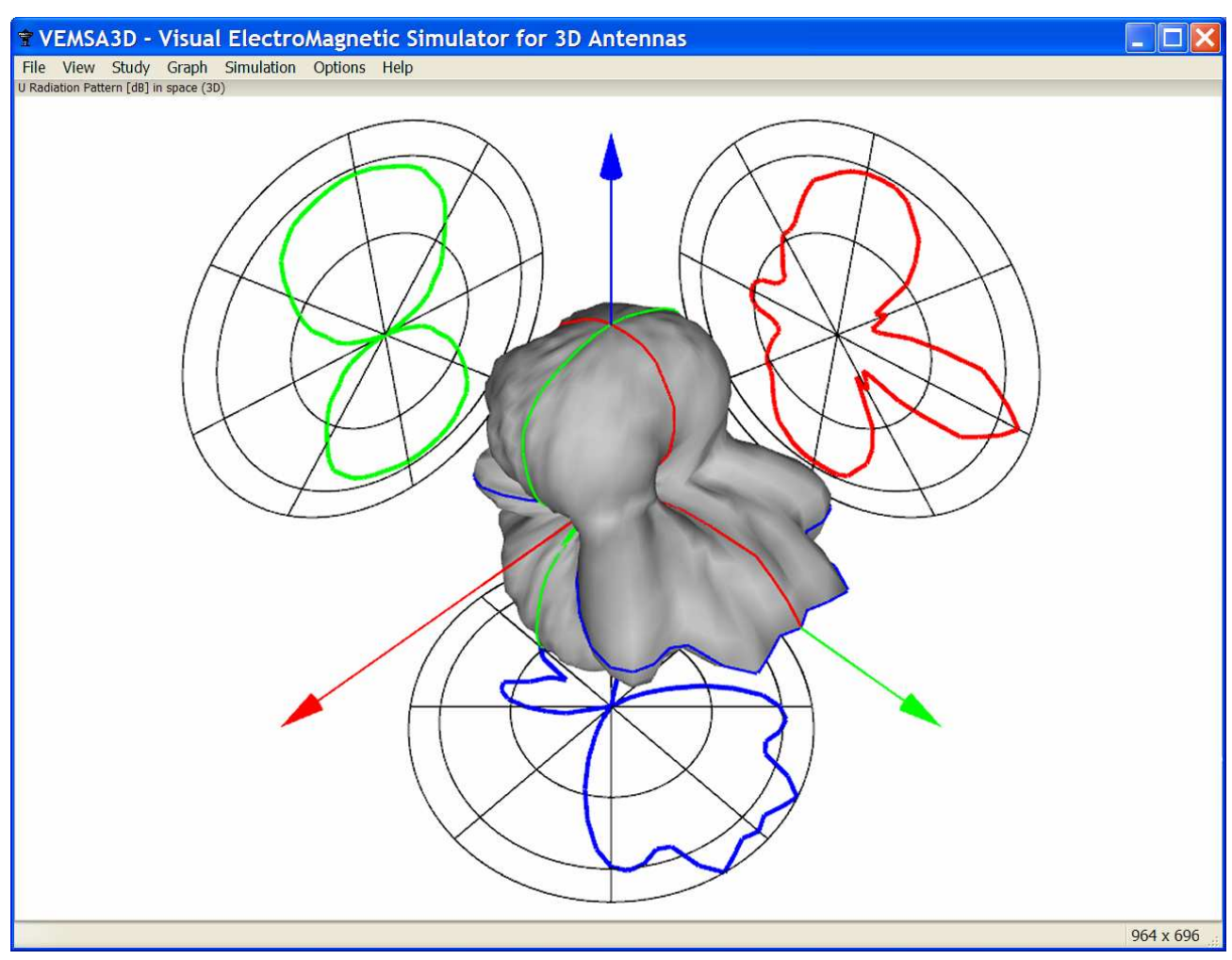

Fig. 16. Horn antenna with bow-tie feeder: Radiation intensity.

## V. CONCLUSION

The first stable version of a visual EM simulator has been developed and released as FLOSS. The computational results of its use were found to be in good agreement with experimental measurements as well with the comparable freeware simulator [4NEC2X]. Since the number of possible modifications and additions to the attributes of [VEMSA3D] seems to be endless, no attempt will be made to suggest a particular direction for its future development: any contribution from the antenna community is very welcomed.